\documentclass[12pt]{article}
\usepackage{amsfonts, amsmath, amsthm, amssymb,graphicx,natbib,gensymb,pdflscape,afterpage,capt-of,lipsum}
\title{The predictive power of the business and bank sentiment of firms: A high-dimensional Granger Causality approach}
\author{Ines Wilms$^a$\footnote{Corresponding author: ines.wilms@kuleuven.be}, Sarah Gelper$^b$ and Christophe Croux$^a$ \\
\footnotesize{$^a$ Faculty of Economics and Business, KU Leuven} \\
\footnotesize{$^b$ Innovation, Technology Entrepreneurship \& Marketing Group, Eindhoven University of Technology}}
\date{ }
\usepackage[left=1in,top=1in,right=1in,bottom=1in,nohead,paperwidth=8.5in, paperheight=11in]{geometry} %1 inch margins, 8 1/2 x 11-inch pages
\begin{document}
\maketitle

\noindent
\textbf{Abstract.}
We study the predictive power of industry-specific economic sentiment indicators for future macro-economic developments. In addition to the sentiment of firms towards their own business situation, we study their sentiment with respect to the banking sector - their main credit providers. The use of industry-specific sentiment indicators results in a high-dimensional forecasting problem. To identify the most predictive industries, we present a bootstrap Granger Causality test based on the Adaptive Lasso. This test is more powerful than the standard Wald test in such high-dimensional settings. Forecast accuracy is improved by using only the most predictive industries rather than all industries.   \\


\newpage
\section{Introduction}
%Business and consumer surveys
Sentiment indicators are often considered to be among the most important leading indicators of the real economy \citep{Dreger13} and are therefore closely followed by business cycle analysts, central banks and business owners (\citealp{Vuchelen04}, \citealp{Claveria07}, \citealp{Martinsen14}).
%Previous literature on consumer sentiment and future spending
However, studies on the predictive power of sentiment indicators find mixed results. %\citep{Claveria07, Taylor07}.
While many studies find that sentiment indicators have predictive power for future economic developments  (\citealp{Kumar95}, \citealp{Hansson05}, \citealp{Lemmens05}, \citealp{Abberger07}, \citealp{Klein10}, \citealp{Christiansen14}), others  conclude that sentiment indicators provide only limited information for predicting economic variables
(\citealp{Cotsomitis06}, \citealp{Claveria07}, \citealp{Dreger13} and \citealp{Bruno14}). 

An important communality between these studies is the use of aggregate sentiment indicators.  This paper, instead, examines the predictive power of disaggregate sentiment indicators. 
Especially in the context of business sentiment -- as is the topic of this paper -- some segments have more predictive power than others. Here, we segment firms according to their industry. Our methodology takes into account that the different industry segments might contain predictive power for different macro-economic indicators. %growth in several industries, including manufacturing, capital goods, consumer goods, wholesale and retail trade.

% Methodology
To study the predictive power, we use a Granger Causality approach. A (set of) time series is said to Granger Cause another time series if the former has incremental predictive power for predicting the latter.
% Granger Causality in LD Settings
Granger Causality tests in \textit{low-dimensional} time series settings have a long history. They are used, among others, 
in macro-economics to study the predictive power of monetary aggregates for output and price variables \citep{Sahoo10}, 
in operational research to study the predictive power of academic literature for practitioner literature \citep{Ghosh10}, 
or in finance to study the predictive power of volume for stock prices \citep{Blasco05}.  
% Granger Causality in HD Settings
Because predictive analysis based on disaggregate sentiment indicators requires handling a large number of such indicators, we introduce a Granger Causality testing procedure applicable to \textit{high-dimensional} time series.

% Recent literature on infernce in high-dimensional models
Recently, a small but growing literature on inference in penalized regression models for \textit{cross-sectional} data has arisen, such as \cite{Wasserman09}, \cite{Meinshausen09} and \cite{Chatterjee11}.
We extend the residual bootstrap procedure  of \cite{Chatterjee11}  to high-dimensional \textit{time series} data. The bootstrap test statistic, based on the Adaptive Lasso \citep{Zou06}, identifies those industry segments whose predictive power is statistically significant.  Our simulation study shows that this test statistic is more powerful than the standard Wald test statistic in a high-dimensional setting. Furthermore, important gains in forecast accuracy are obtained by not using all industry segments but by first selecting the most predictive ones using the bootstrap test statistic.

%Limitation : no bank sentiment data
We use a unique data set that not only measures the sentiment of firms towards their own situation  (``\textit{business} sentiment") -- as is classical for sentiment indicators -- but also measures the sentiment of firms towards the banking industry (``\textit{bank} sentiment"). For the economy to be able to grow, it is essential that firms have access to credit, typically provided by banks. Especially in the aftermath of the recent economic downturn and banking crises, distressed banks can constrain the economy (\citealp{Kroszner07}, \citealp{DellAriccia08}, \citealp{Fernandez13}). 
To the best of our knowledge, we are the first to study the importance of sentiment towards the banking industry.

%Structure of paper:
The remainder of this article is structured as follows. Section \ref{data} describes the data on the business and bank sentiment, as well as the macro-economic indicators. Section \ref{GC} introduces Granger Causality Testing in high-dimensional time series models. In Section \ref{simulation}, a simulation study shows the good performance of our methodology in terms of size and power of the test statistic and forecast accuracy. 
In Section \ref{application}, we apply the proposed methodology to identify the most predictive industry segments for several future macro-economic indicators. In Section \ref{forecast}, we show that forecast accuracy can be improved by using only the most predictive industry segments instead of all industry segments.  Finally, Section \ref{conclusion} concludes. 

\section{Data \label{data}}
We use a unique data set  provided to us by EUWIFO, the European Economic Research Institute. EUWIFO is an owner-managed business that conducts business climate interviews.
By conducting interviews with firms spread over Germany, EUWIFO gathers information on the confidence these firms have in their own economic situation and in the banking sector. Firms are divided into segments  according to  the industry in which they are active based on their NACE code. These 10 industry segments are listed in Table \ref{businesses}.

\begin{table}
\footnotesize
\caption{Industry Segments. Businesses are divided into 10 industry segments. \label{businesses}}
\resizebox{0.74\textwidth}{!}{\begin{minipage}{\textwidth}
\centering
\begin{tabular}{lllll} \hline
\textbf{Industry} & && \textbf{Description} & \textbf{Sector}  \\ \hline
Industry 1 	&  && \small{Agriculture, forestry, fishing, mining and quarrying and other industry } 		& Primary	 \\
Industry 2	&  && \small{Manufacturing}				& Secondary				\\
Industry 3 	&  && \small{Construction} 				& Secondary				 \\
Industry 4	&  && \small{Wholesale and retail trade, transportation and storage accomodation and food and service activities } & Tertiary \\	
Industry 5 	&  && \small{Information and communication}	& Quaternary			\\
Industry 6 	&  && \small{Financial and insurance activities} & Quaternary	\\
Industry 7 	&  && \small{Real estate activities} 	& Quaternary				\\
Industry 8 	&  && \small{Professional, scientific, technical administration and support service activities}& Quaternary	\\
Industry 9 	&  && \small{Public administration, defence, education, } & Quaternary  \\			
Industry 10 &  && \small{Other services} 	& Quaternary	\\
\hline
\end{tabular}
\end{minipage} }
\end{table}

The interviews consist of two parts. In the first part, the Business Survey, firms are asked to assess their own situation. In the second part, the Bank Survey, firms are asked to assess the German bank sector.

\paragraph{Business Survey} Each firm receives 9 questions to assess their own economic situation. They are asked to assess changes (this year compared to last year) in  (1) turnover,  (2) earnings,  (3) number of employees, (4) investments,  (5) incoming domestic orders,  (6) incoming foreign orders, (7) utility and maintenance costs,  (8) tax burden, and  (9) cost through government red tape. 
For each question, answers are favorable, neutral or unfavorable. For all the firms within an industry segment, a balance of opinion indicator is calculated for each question, being the percentage of favorable answers minus the percentage of unfavorable answers. As we construct 9 sentiment indicators for each of the 10 industries, this amounts to 90 business sentiment indicators. 

\paragraph{Bank Survey} Each firm is asked to assess the German bank sector. In total, 243 German banks are included in the Bank Survey. Each firm first has to indicate which of these 243 German banks they know. For the banks they know, they are asked to assess their \textit{consideration} towards that specific bank and the \textit{reputation} of that specific bank.
Answers are either favorable or unfavorable and a balance of opinion indicator is calculated for each question.
We include three indicators: the average consideration indicator, averaged over all German banks,  the consideration  indicator towards the Sparkassen, and the consideration indicator towards the Volksbanken. The latter two are the most well known banks in Germany.  We also construct three reputation indicators per industry segment following an analogous approach. As we construct three bank consideration and three bank reputation indicators for each of the 10 industries, this amounts to 60 bank sentiment indicators. 

\medskip

Joining the 90 business sentiment indicators and the 60 bank sentiment indicators results in a total of 150 time series. We combine all 150 sentiment indicators in one high-dimensional data set.  All time series are observed over $T=40$ months (January 2012-April 2015). 
We study the predictive power of these sentiment indicators for 8 German macro-economic indicators (Table \ref{Responses}).

\begin{table}
\caption{Macro-economic indicators. All time series are seasonally adjusted (Eurostat).\label{Responses}}
\resizebox{0.63\textwidth}{!}{\begin{minipage}{\textwidth}
\centering
\begin{tabular}{llll} \hline
\textbf{Indicator} 	&&& \textbf{Description} \\ \hline
IP-A1		&&& Production in industry: Mining and quarrying; manufacturing; electricity, gas, steam and air conditioning supply \\
IP-A2		&&& Production in industry: Construction, Mining and quarrying; manufacturing; electricity, gas, steam and air conditioning supply\\
IP-M 		&&& Production in industry: Manufacturing \\
IP-E		&&& Production in industry: Energy\\
IP-CaGo		&&& Production in industry: Capital goods \\
IP-CoGo		&&& Production in industry: Consumer goods \\
RT		&&& Retail Trade, except of motor vehicles and motorcycles \\
WS		&&& Wholesale Trade, except of motor vehicles and motorcycles \\ \hline
\end{tabular}
\end{minipage} }
\end{table}

The 150 time series are grouped into blocks by industry segment (cfr. Table \ref{businesses}). 
%The aim is to identify the most predictive industry blocks for future macro-economic developments. 
For each industry segment, we have one block of 9 indicators from the Business Survey and one block of 6 indicators from the Bank Survey. Our methodology is such that we select either all 9 business sentiment indicators for an industry, or none. Similarly, we will select either all 6 bank sentiment indicators for an industry or none. 
This way, we can investigate the difference in predictive power between the business and bank sentiment indicators for the 10 industries.
To identify the most predictive blocks, we perform joint hypothesis tests. We  test if the set of indicators in a particular block 
Granger Causes a particular macro-economic indicator.   %A set of time series is said to Granger Cause another time series if the former has incremental predictive power for predicting the latter.
This predictive analysis involves a large number of disaggregate sentiment indicators. 
In the next section, we introduce a  Granger Causality testing procedure that can handle such a high-dimensional situation.

\section{High-dimensional Granger Causality Testing \label{GC}}
Performing Granger Causality tests on a data set with many time series relative to the length of the series is challenging.
% Challenge
In these high-dimensional settings,
estimation by standard  procedures becomes inaccurate. In our sentiment application, the number of time series (i.e.\ $k=150$) even exceeds the length of the time series (i.e.\ 40), making it impossible to use standard estimation procedures.
Penalized estimation brings an outcome. 

\subsection{Penalized Maximum Likelihood estimation\label{PVAR}}
Let ${y}_t$ be a one-dimensional stationary time series. We assume that ${y}_t$ follows a ARX($p$) model, i.e.\ an autoregressive model of order $p$ with $k$ predictor time series collected in the $(k \times 1)$ vector ${\bf x}_t$:
\begin{equation}\label{varp}
{ y}_t = {b}_1 {y}_{t-1}  + {b}_2 {y}_{t-2}  + \ldots + {b}_p {y}_{t-p} + {\bf a}_1 {\bf x}_{t-1}  + {\bf a}_2 {\bf x}_{t-2}  + \ldots + {\bf a}_p {\bf x}_{t-p} + {e}_t \, ,
\end{equation}
where ${b}_1$ to ${b}_p$ are the autoregressive parameters, the parameters ${\bf a}_1$ to ${\bf a}_p$ are $(1 \times k)$ vectors and the error term ${e}_t$ is assumed to follow a $N({0},\sigma)$ distribution. %Let $\boldsymbol \Sigma^{-1}= \boldsymbol \Omega$.
We assume, without loss of generality, that all time series are mean centered such that no intercept is included.

If the number of components in  ${\bf x}_t$ is large,  the number of unknown parameters  in equation \eqref{varp} explodes.
To ensure accurate estimation, we use Penalized Maximum Likelihood estimation (e.g. \citealp{Zou06} in a regression context, or \citealp{Gelper15} in a time series context).
Write model \eqref{varp} in matrix notation as
\begin{equation}\label{stacked}
 {\bf y} = {\bf X} \boldsymbol \beta + {\bf e} \, ,
\end{equation}
where 
$ {\bf y}$ is the column vector $(y_1,\ldots,y_T)$, and
the matrix ${\bf X} = (\underline{{\bf Y}}_1, \ldots, \underline{{\bf Y}}_p, \underline{{\bf X}}_1, \ldots, \underline{{\bf X}}_p)$. %, is of dimension $(n q \times p q^2)$.
Here $\underline{{\bf Y}}_j$ is  $(T \times  1)$, containing the values of the time series at lag $j$ in its column; and  $\underline{{\bf X}}_j$ is an $(T \times  k)$ matrix, containing the values of the $k$  predictor time series at lag $j$ in its columns,  for $1 \leq j \leq p$. 
The vector $\boldsymbol \beta$ contains the parameters values $b_1,\ldots,b_p,{\bf a}_1, \ldots,{\bf a}_p$, and has length $p(1+k)$. In case $p(1+k)>T$, the Maximum Likelihood estimator does not exist. The Penalized Maximum Likelihood estimator is, however, still computable.

The penalized estimator of the regression parameter $\boldsymbol\beta$
is obtained  by minimizing the negative log likelihood with a penalization on the elements of $\boldsymbol\beta$:
\begin{equation}\label{mincrit}
\widehat{\boldsymbol\beta}_{\lambda} = \underset{\boldsymbol\beta}{\operatorname{argmin}} \, \frac{1}{T} ({\bf y}-{\bf X} \boldsymbol \beta)^{\prime} ({\bf y}-{\bf X} \boldsymbol \beta)   + \lambda \sum_{i=1}^{p(1 + k)} \hat{w}_i |\beta_i| \, ,
\end{equation}
where $\hat{w}_i$ are weights and $\lambda>0$ is a sparsity parameter. This estimator is the Adaptive Lasso \citep{Zou06}. It generalizes the popular Lasso (e.g. \citealp{Hastie09}, Chapter 3) which shows good performance in operational research (e.g. \citealp{Ballings15}, \citealp{Huang14}). The Adaptive Lasso  ensures that the bootstrap (Section \ref{teststat}) is consistent  \citep{Chatterjee11}.  We take the weights of the Adaptive Lasso  $\hat{w}_i=1/|\hat{\beta}_i^{\text{ridge}}|$, where the Ridge estimator (\citealp{Hastie09}, Chapter 3) is 
\begin{equation}
\hat{\boldsymbol\beta}^{\text{ridge}}_{\lambda} = \underset{\boldsymbol\beta}{\operatorname{argmin}} \frac{1}{T} ({\bf y}-{\bf X} \boldsymbol\beta)^{\prime} ({\bf y}-{\bf X} \boldsymbol\beta) + \lambda_{\text{ridge}} \sum_{i=1}^{p(1 + k)}  \beta^2_i. \nonumber
\end{equation}

The sparsity parameter $\lambda$ and the order of the ARX, $p$, are selected using the Bayesian Information Criterion (BIC) (e.g. \citealp{Abegaz13} and references therein):
\begin{equation}\label{eq: BICbetaB}
\text{BIC}_{\lambda} = T \cdot\text{log}\left(\frac{1}{T} ({\bf y}-{\bf X} \widehat{\boldsymbol \beta}_{\lambda})^{\prime} ({\bf y}-{\bf X} \widehat{\boldsymbol \beta}_{\lambda}) \right) + df_{\lambda} \cdot \log(T), \nonumber
\end{equation}
where $df_{\lambda} $ equals the number of non-zero estimated regression coefficients. 
We solve  \eqref{mincrit} over a range of values for $\lambda$ and select the one with lowest value of the BIC.
To select the order of the ARX model, we estimate the ARX model for different values of $p$, each time using the optimal value of  $\lambda$ for that value of $p$. We then select the order $p$ of the ARX model again by minimizing the BIC.

\subsection{Granger Causality in the ARX framework \label{GC_ARX}}
% Granger Causality
We partition the vector ${\bf x}_t$ in different blocks, and denote the $j^{th}$ block of ${\bf x}_t$ by ${\bf x}_{t,j}$, consisting of $k_j$ time series. In the ARX model \eqref{varp}, denote the $j^{th}$ block of coefficients at lag $i$ corresponding to ${\bf x}_{t,j}$ by ${\bf a}_{i,j}$.  
The  multivariate time series ${\bf x}_{t,j}$ is said to Granger Cause ${y}_t$ if the former has incremental predictive power for  the latter.
We say that ${\bf x}_{t,j}$ does not Granger Cause ${y}_t$ if the coefficients on all lags of ${\bf x}_{t,j}$ are equal to zero, i.e.\ ${\bf a}_{1,j}=\ldots={\bf a}_{p,j}={\bf 0}$. %, where ${\bf A}_{i,j}$ contains the $k_j$ coefficients at lag $i$ corresponding to  multivariate time series ${\bf x}_{t,j}$.

%Granger Lasso Inference vs Granger Lasso Selection
%Granger Causality in high-dimensional time series models has recently gained some attention in the literature. 
The Adaptive Lasso estimator in \eqref{mincrit} is sparse, meaning that some of its elements are exactly zero. The larger the value of $\lambda$, the sparser the estimator.
The ``Granger Lasso Selection" method (e.g. \citealp{Fujita07}, \citealp{Bahadori13}) says that a time series ${\bf x}_{t,j}$  Granger Causes ${y}_t$ if at least one of the corresponding parameters  ${\bf a}_{1,j},\ldots,{\bf a}_{p,j}$ is estimated as non-zero. 
%They estimate the models using the Lasso (\citealp{Hastie09}, Chapter 3). As a result, a sparse solution is obtained: some of the parameters are estimated as exactly zero. These ``Granger Lasso Selection" methods then infer the Granger Causality relations directly from the sparsity of the parameter estimates. 
%
Our approach is different, we infer Granger Causality relations from a bootstrap testing procedure.
 
\subsection{Granger Lasso test \label{teststat}}
The null hypothesis that a block of time series ${\bf x}_{t,j}$ is not Granger Causing ${y}_t$ can be stated as
\begin{equation}
H_{0}: \ {\bf R}_j\boldsymbol \beta = \bf 0,
\end{equation}
where ${\bf R}_j$ is a suitable $pk_j \times p(1+k)$ matrix. The elements of ${\bf R}_j$ are either zero or one. We assign the value one to the elements of ${\bf R}_j$ corresponding  to the autoregressive parameters ${\bf a}_{1,j},\ldots,{\bf a}_{p,j}$. The corresponding Wald test statistic is given by 
\begin{equation} \label{eqteststat}
Q = ({\bf R}_j \widehat{\boldsymbol\beta})^{\prime} ({\bf R}_j \text{Cov}(\widehat{\boldsymbol\beta}) {\bf R}_j^{\prime})^{-1} ({\bf R}_j \widehat{\boldsymbol\beta}).
\end{equation}

To bootstrap this test statistic, we use the following residual bootstrap procedure (\citealp{Rao12}):
\begin{enumerate}
\item Estimate the model under the null hypothesis, i.e.\ model \eqref{varp} with the block ${\bf x}_{t,j}$ removed at the right-hand-side. Compute the centered residuals $\widehat{\varepsilon}_{t}, \text{for} \ t=1,\ldots,T$.
\item Let $B=500$ be the number of bootstraps. For $b=1,\ldots, B$:
\begin{enumerate}
\item[(a)] Construct the bootstrap time series $y_t^*$ from model \eqref{varp} with the parameter estimates from  step 1 and with bootstrap errors  $\varepsilon_{t}^{*} =  \widehat{\varepsilon}_{\mathcal{U}_{t}} $ with $\mathcal{U}_{t}, t=1, \ldots, T$ an i.i.d. sequence of discrete random variables uniformly distributed on $\{1,\ldots, T\}$. The predictor time series are kept fixed.
\item[(b)] Apply the Penalized Maximum Likelihood estimator of equation \eqref{mincrit} to the bootstrap sample.  Denote the  bootstrap estimate  by $\widehat{\boldsymbol\beta}^*_b$.
\item[(c)] Compute the bootstrap statistic $Q^{*}_b=({\bf R}_j \widehat{\boldsymbol\beta}^*_b)^{\prime} ({\bf R}_j \text{Cov}(\widehat{\boldsymbol\beta}) {\bf R}_j^{\prime})^{-1} ({\bf R}_j \widehat{\boldsymbol\beta}^*_b)$.
\end{enumerate}
\item Compute $$ \text{mid $p$-value }= \frac{1}{B}  \sum_{b=1}^{B} \Big ( I(Q^{*}_{b} >   Q)  + \frac{1}{2} I(Q^{*}_{b} =   Q) \Big ),$$
 with $Q^{*}_{b} \ (\text{for } \ b=1, \ldots, B)$ $B$ independent bootstrap statistics. $I(\cdot)$ is an indicator function that takes on the value one if its argument is true and equals zero otherwise. We use the mid $p$-value \citep{Lancaster49} since it may occur that the value of the test statistic and the bootstrap test statistic are both equal to zero. 
\end{enumerate}

\section{Simulation study \label{simulation}}
By means of a simulation experiment, we (i) evaluate the size and power of the Granger Lasso test and  (ii) conduct a forecast exercise.
We generate ${y}_t$ according to the following ARX(1) model
\begin{equation}
{ y}_t = 0.5{y}_{t-1}  + {\bf a}_1{\bf x}_{t-1}+ {e}_t, 
\end{equation}
where  ${e}_t \sim N(0,0.1)$. The predictors are generated as autoregressive processes ${\bf x}_t = {\bf C}{\bf x}_{t-1} + {\bf u}_t,$ with ${\bf u}_t \sim N_k({\bf 0},0.1 {\bf I})$, ${\bf C}=0.5{\bf I}$ and ${\bf I}$ the $k$-dimensional identity matrix. The model parameters are chosen according to the four designs detailed in Table \ref{SimDGP}.  The first three designs are the same except for the number of time series $k$. In design two and three, we add more non-informative time series to the model, i.e.\ time series with a coefficient equal to zero.
The standard Maximum Likelihood estimator is computable in these three designs. The last design corresponds to the design of our sentiment application, with $k=150$ predictor time series and $T=40$. Here, only the Penalized Maximum Likelihood estimator is computable.

\begin{table}
\caption{Simulation designs. \label{SimDGP}}
\resizebox{0.7\textwidth}{!}{\begin{minipage}{\textwidth}
\centering
\begin{tabular}{lllll} \hline
Design  && under $H_0$ && under $H_A$ \\  \hline
&&&& \\
$T=100,k=25$  && ${\bf a}_1= \begin{bmatrix} {\bf 0.2}_{1 \times 5} & {\bf 0}_{1 \times 5} & {\bf 0}_{1 \times 5} & {\bf 0}_{1 \times (k-15)}
\end{bmatrix}$ &&  ${\bf a}_1= \begin{bmatrix} {\bf 0.2}_{1 \times 5}  & {\bf 0.2}_{1 \times 5} & {\bf 0}_{1 \times 5} & {\bf 0}_{1 \times (k-15)} \end{bmatrix}$\\
$T=100,k=50$ && ${\bf a}_1= \begin{bmatrix} {\bf 0.2}_{1 \times 5} & {\bf 0}_{1 \times 5} & {\bf 0}_{1 \times 5} & {\bf 0}_{1 \times (k-15)}
\end{bmatrix}$ &&  ${\bf a}_1= \begin{bmatrix} {\bf 0.2}_{1 \times 5}  & {\bf 0.2}_{1 \times 5} & {\bf 0}_{1 \times 5} & {\bf 0}_{1 \times (k-15)} \end{bmatrix}$\\
$T=100,k=75$ && ${\bf a}_1= \begin{bmatrix} {\bf 0.2}_{1 \times 5} & {\bf 0}_{1 \times 5} & {\bf 0}_{1 \times 5} & {\bf 0}_{1 \times (k-15)}
\end{bmatrix}$ &&  ${\bf a}_1= \begin{bmatrix} {\bf 0.2}_{1 \times 5}  & {\bf 0.2}_{1 \times 5} & {\bf 0}_{1 \times 5} & {\bf 0}_{1 \times (k-15)} \end{bmatrix}$\\
$T=40,k=150$ && ${\bf a}_1= \begin{bmatrix} {\bf 0.4}_{1 \times 9}  &  {\bf 0}_{1 \times 9} &  \ldots &   {\bf 0}_{1 \times 9} &  {\bf 0}_{1 \times 6} &  \ldots &  {\bf 0}_{1 \times 6}
\end{bmatrix}$ &&  ${\bf a}_1= \begin{bmatrix} {\bf 0.4}_{1 \times 9}  &  {\bf 0.4}_{1 \times 9} &   {\bf 0}_{1 \times 9} &\ldots &   {\bf 0}_{1 \times 9} &  {\bf 0}_{1 \times 6} &  \ldots &  {\bf 0}_{1 \times 6}
\end{bmatrix}$\\
%ARX $T=40,k=150$  && $p=1$ && $H_0: \beta_{10}=\ldots=\beta_{18}=0$ && ${ y}_t = 0.5{y}_{t-1}  + {\bf A}_1{\bf x}_{t-1}+ {e}_t$&& \\
%%&  &&  && ${\bf e}_t \sim N(0,0.1)$ && \\
%&& $T=40$ &  &&& ${\bf A}_1= \begin{bmatrix} {\bf 0.4}_{1 \times 9} & {\bf 0}_{1 \times (k-9)}
%\end{bmatrix}$&& ${\bf A}_1= \begin{bmatrix} {\bf 0.4}_{1 \times 9}  & {\bf 0.4}_{1 \times 9}& {\bf 0}_{1 \times (k-18)} \end{bmatrix}$\\
%&& $k=150$ &  &&& ${\bf e}_t$  and ${\bf x}_t$ generated as before   &&  \\
 \hline
\end{tabular}
\end{minipage} }
\end{table}

%\begin{table}
%\caption{Simulation designs. \label{SimDGP}}
%\resizebox{0.67\textwidth}{!}{\begin{minipage}{\textwidth}
%\centering
%\begin{tabular}{lllllllll} \hline
%Design && Dimensions 				  && Tested Null Hypothesis  && DGP under $H_0$ && DGP under $H_A$ \\  \hline
%ARX $T=100,k=25$  && $p=1$ && $H_0: \beta_7=\ldots=\beta_{11}=0$ && ${ y}_t = 0.5{y}_{t-1}  + {\bf A}_1{\bf x}_{t-1}+ {e}_t$&& \\
% &&  $T=100$&&  && ${\bf e}_t \sim N(0,0.1)$ && \\
%  &&  $k=25$  &  &&& ${\bf A}_1= \begin{bmatrix} {\bf 0.2}_{1 \times 5} & {\bf 0}_{1 \times (k-5)}
%\end{bmatrix}$&& ${\bf A}_1= \begin{bmatrix} {\bf 0.2}_{1 \times 5}  & {\bf 0.2}_{1 \times 5}& {\bf 0}_{1 \times (k-10)} \end{bmatrix}$\\
%  &&  &  &&& ${\bf x}_t = {\bf C}{\bf x}_{t-1} + {\bf u}_t$ &&  \\
% && &  &&& ${\bf u}_t \sim N_k({\bf 0},0.1 {\bf I})$ &&  \\
% && &  &&& ${\bf C}=0.5{\bf I}$ &&  \\ 
%
%ARX $T=100,k=50$ && \multicolumn{7}{l}{First design, except for $k=50$} \\ 
%ARX $T=100,k=75$ && \multicolumn{7}{l}{First design, except for $k=75$} \\
%ARX $T=40,k=150$  && $p=1$ && $H_0: \beta_{10}=\ldots=\beta_{18}=0$ && ${ y}_t = 0.5{y}_{t-1}  + {\bf A}_1{\bf x}_{t-1}+ {e}_t$&& \\
%%&  &&  && ${\bf e}_t \sim N(0,0.1)$ && \\
%&& $T=40$ &  &&& ${\bf A}_1= \begin{bmatrix} {\bf 0.4}_{1 \times 9} & {\bf 0}_{1 \times (k-9)}
%\end{bmatrix}$&& ${\bf A}_1= \begin{bmatrix} {\bf 0.4}_{1 \times 9}  & {\bf 0.4}_{1 \times 9}& {\bf 0}_{1 \times (k-18)} \end{bmatrix}$\\
%&& $k=150$ &  &&& ${\bf e}_t$  and ${\bf x}_t$ generated as before   &&  \\
% \hline
%\end{tabular}
%\end{minipage} }
%\end{table}

For each design, we consider a data generating process under the null hypothesis $H_0$ and under the alternative hypothesis $H_A$.
We divide the time series ${\bf x}_t$ and the corresponding coefficient vector ${\bf a}_1$ into several blocks, as can be seen from Table \ref{SimDGP}. The first block of time series Granger Cause the response both  under $H_0$ and  under $H_A$. The  second block of time series Granger Cause the response only under $H_A$. The remaining blocks of time series never Granger Cause the response. 
In the first three designs,  block one to three each contain five time series, the fourth block contains the remaining ones. In the last design, there are 20 blocks, similar to our sentiment application.

\subsection{Size and power of the test statistic \label{sizepower}}
We test the null hypothesis that the second block of time series does not Granger Cause the response. 
We compare the performance of Granger Lasso test to the  standard Wald test computed from the standard Maximum Likelihood (ML) estimator. %, and (2) the ``Granger Lasso Selection" procedure  (cfr. Section \ref{PVAR}).
\medskip

% Size and Power of bootstrap test statistic
To study the \textit{size} of the test statistic, we simulate $N=1000$ time series under the null hypothesis and compute the simulated size, i.e.\ the proportion of simulation runs were the null hypothesis is rejected: 
\begin{equation}
\text{Simulated size} = \frac{1}{N} \sum_{j=1}^{N} I(p^{H_0}_j < \alpha),
\end{equation}
where
$p^{H_0}_j$ is the mid $p$-value obtained in simulation run $j=1,\ldots,N$, and $\alpha$ is the pre-specified significance level. We consider $\alpha=0.01$ and $\alpha=0.05$.
\smallskip

\textit{Results.}
Table \ref{sizeARX} shows the simulated sizes for the standard Wald test and the Granger Lasso test. % and  Granger Lasso Selection procedure. 
The simulated sizes of  the Granger Lasso test  and the standard Wald test are both close to the nominal size $\alpha$ in the  design with $T=100, k=25$. When the number of time series increases relative to the length of the time series (i.e.\ second and third design), the Granger Lasso test  remains accurately sized whereas the standard Wald test statistic gets distorted: its simulated size deviates strongly from the nominal size. %The Granger Lasso Selection procedure has a much higher false rejection rate than the other two procedures.
% HD: false rejection rate
In the last design, only the Granger Lasso test is available.
For both $\alpha=0.01$ and $\alpha=0.05$,  the Granger Lasso test  is reasonably accurately sized. %Granger Lasso Inference again has a  much lower false rejection rate than  Granger Lasso Selection.

%\begin{table}
%\caption{False rejection rates for the standard Wald test, the Granger Lasso Inference test  and the  Granger Lasso Selection procedure. \label{sizeARX} }
%\resizebox{0.82\textwidth}{!}{\begin{minipage}{\textwidth}
%\centering
%\begin{tabular}{llcccccccccc} \hline
%Design &&& \multicolumn{2}{c}{Standard Wald Test} &&& \multicolumn{2}{c}{Granger Lasso Inference} && \multicolumn{2}{c}{Granger Lasso Selection}\\
%		&&&$\alpha=0.01$ & $\alpha=0.05$  &&& $\alpha=0.01$ & $\alpha=0.05$  && &\\ \hline
%
%ARX $T=100, k=25$					&&&  0.017 &  	0.064  	&&& 0.013 & 0.058	 	&&&  0.730\\ 
%ARX $T=100, k=50$					&&&  0.025 & 0.079 	   	&&& 0.010 &	0.052  &&& 0.649\\ 
%ARX $T=100, k=75$					&&&  0.035 & 0.082  	   	&&& 0.015 & 0.051 	  &&& 0.611 \\ 
%ARX $T=40, k=150$					&&& NA 		&  NA 		&&&  0.007 	& 0.051  &&& 0.717  \\ \hline	
%\end{tabular}
%\end{minipage} }
%\end{table}

\begin{table}
\small
\caption{Simulated sizes for the  Wald test and Granger Lasso test. \label{sizeARX} }
\resizebox{1\textwidth}{!}{\begin{minipage}{\textwidth}
\centering
\begin{tabular}{llccccccc} \hline
Simulation design &&& \multicolumn{2}{c}{Wald test} &&& \multicolumn{2}{c}{Granger Lasso test} \\
		&&&$\alpha=0.01$ & $\alpha=0.05$  &&& $\alpha=0.01$ & $\alpha=0.05$  \\ \hline

$T=100, k=25$					&&&  0.017 &  	0.064  	&&& 0.013 & 0.058	 \\ 
$T=100, k=50$					&&&  0.025 & 0.079 	   	&&& 0.010 &	0.052  \\ 
$T=100, k=75$					&&&  0.035 & 0.082  	   	&&& 0.015 & 0.051 	 \\ 
$T=40, k=150$					&&& NA 		&  NA 		&&&  0.007 	& 0.051  \\ \hline	
\end{tabular}
\end{minipage} }
\end{table}

\bigskip

To study the \textit{power} of the test statistic, we use size-power curves (see \citealp{Davidson98}). Size-power curves are constructed using two empirical distribution functions. %(EDFs): one for an experiment under the null hypothesis, and one for an experiment under the alternative hypothesis.
We carry out the following steps:
\begin{enumerate}
\item Simulate $N=1000$ time series under the null hypothesis. Compute for each simulation run $j=1,\ldots,N$ the mid $p$-value $p^{H_0}_j$. Calculate the empirical distribution function of the $p$-values:
\begin{equation}
\widehat{F}^{H_0}(x_i) = \frac{1}{N} \sum_{j=1}^{N} I(p^{H_0}_j \leq x_i), \nonumber
\end{equation}
for a grid of values $x_i, i=1,\ldots,m$ between zero and one.
%where $I(\cdot)$ is an indicator function that takes on the value one if its argument is true and equals 0 otherwise.
%where we evaluate the EDF at a  grid of values $x_i, i=1,\ldots,m$.
%\begin{equation}
%x_1=0.001,0.002,\ldots,0.010,0.015,\ldots,0.990,0.991,\ldots,x_m=0.999 \text{\ \ \ }(m=215). \nonumber
%\end{equation}
%Extra points in the tails are taken in order not to miss any unusual behavior in the tails \citep{Davidson98}.
\item Simulate $N=1000$ time series under the alternative hypothesis. %The DGPs under the alternative hypothesis are in Table \ref{SimDGP}.
Compute for each simulation run $j=1,\ldots,N$ the mid $p$-value $p^{H_A}_j$. Calculate %the EDF of the $p$-values:
\begin{equation}
\widehat{F}^{H_A}(x_i) = \frac{1}{N} \sum_{j=1}^{N} I(p^{H_A}_j \leq x_i). \nonumber
\end{equation}
\item Plot $\widehat{F}^{H_0}(x_i)$ against $\widehat{F}^{H_A}(x_i)$, for $x_i, i=1,\ldots,m$.
\end{enumerate}

\smallskip

\begin{figure}
\begin{center}
\includegraphics[width=17cm]{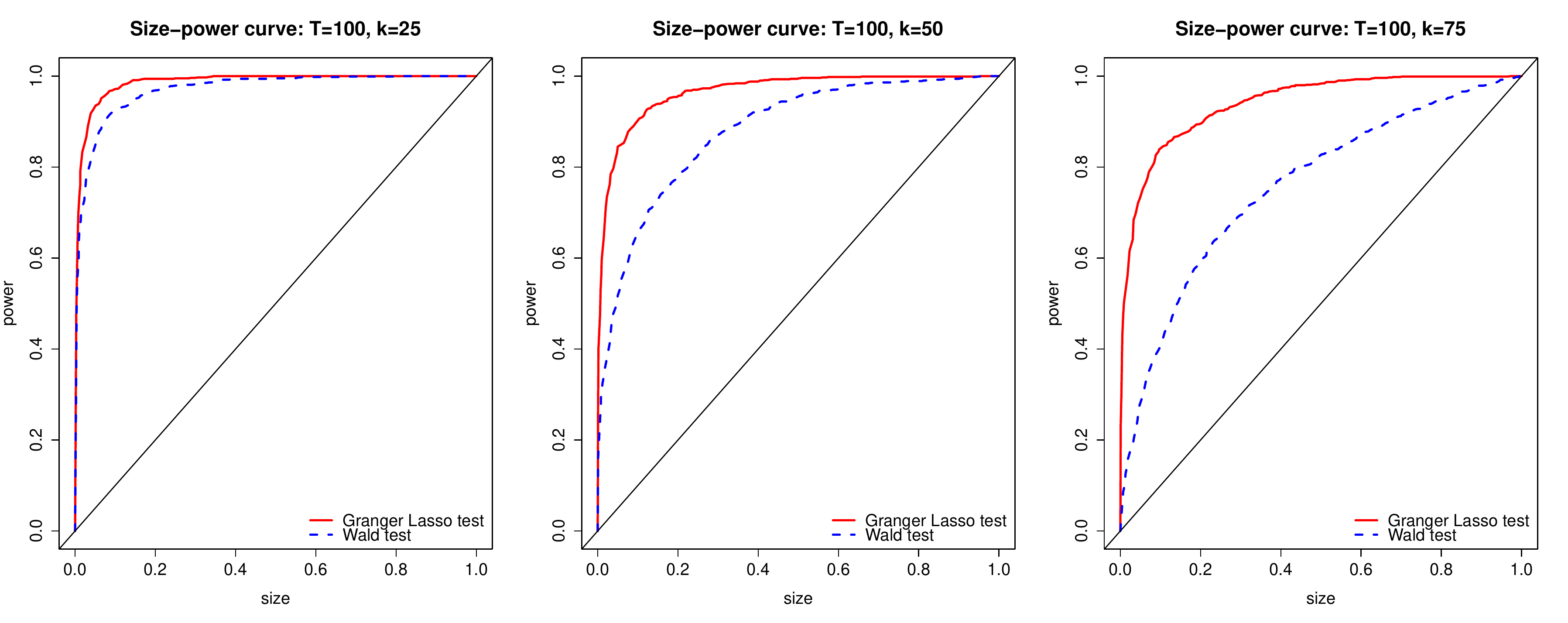}
\caption{Size-power curve of  the Granger Lasso test (solid gray line) and the standard Wald test (dashed line), for increasing number of time series $k=25$ (left), $k=50$ (middle) and $k=75$ (right) with time series length $T=100$. The 45\degree line is indicated as well. \label{SizePowerARX}}
\end{center}
\end{figure}

\textit{Results.}
% LD: size-power curves
Size-power curves of the Granger Lasso test and standard Wald test  are shown in Figure \ref{SizePowerARX} (first three designs). The larger the difference between the size-power curve and the $45$\degree line, the more power the test has. 
For $k=25$ (i.e.\ left panel) both curves are rapidly increasing and very similar. When the number of time series increases (i.e.\ middle and right panel), the size-power curve of the Granger Lasso test  is hardly affected, and achieves a much larger power  than the standard Wald test. %The higher the ratio of the number of time series to the length of the time series, the worse the performance of the standard Wald test since its size-power curve moves towards the $45$\degree line.

\subsection{Forecast exercise \label{SimForecast}}
For forecasting the time series $y_t$, we use a two-step procedure. First, we select predictor time series. Second, we estimate the model with only the selected predictor time series. We consider four selection and four estimation techniques, yielding 16 selection-estimation combinations. We investigate the performance of each combination in forecasting the response.

As selection techniques we consider:
(1) use all time series,
(2) use the standard Wald test  to discard blocks of time series that are not Granger Causing the response,
(3) use Granger Lasso Selection (cfr. Section \ref{PVAR}) to discard blocks of time series that are not Granger Causing the response,  
(4) use the Granger Lasso test to discard blocks of time series that are not Granger Causing the response. Selection technique (4) is our proposed selection technique.
The tests are carried out at a 1\% significance level.
%We use these selection technique to test the null hypothesis that a specific block of time series is not Granger Causing the response. In the first three designs, 4 hypotheses are tested. In the last design, 20 hypotheses are tested (cfr. blocks in Table \ref{SimDGP}).
  
%In the  ARX $T=100, k=25 \ \text{or} \ k=50, \ \text{or} \ k=75$ designs, we divide the time series into four blocks and test the corresponding four null hypotheses. Block one to three each contain five consecutive time series, the fourth block contains the remaining time series.
%%In the  ARX $T=40, k=150$ design, we divide the time series into 20 blocks and  test the corresponding 20 null hypotheses. Block 1 to 10 each consists of 9 consecutive time series, Block 11 to 20 each consists of 6 consecutive time series. This setting corresponds to the type of hypotheses tested in our sentiment application. 

After selecting the predictor time series,  we forecast the response using  either
(1) Maximum Likelihood,
(2) the Adaptive Lasso estimator,
%(3) PML using the Ridge penalty (``Ridge"),
(3) Bayesian shrinkage with the Minnesota prior \citep{Litterman86},
(4)  the Factor Model of \cite{Stock02}.
These are all leading methods for macro-economic forecasting \citep{Inoue08}.
Methods (2) and (3) perform shrinkage. Where the Adaptive Lasso puts some of the estimated coefficients exactly to zero, the Bayesian estimator only shrinks the estimated coefficients towards zero. 
Factor Models reduce the dimension of the predictor time series by extracting a small number of common factors  using principal component analysis.\footnote{The number of factors $r$ is determined by calculating the maximum eigenvalue ratio criterion $\hat{r}_j = \hat{\lambda}_j/\hat{\lambda}_{j+1}$ for $j=1,\ldots,\text{k}-1$ from the eigenvalues $\hat{\lambda}_j,\ldots,\hat{\lambda}_{\text{k}}$ and selecting $r=\text{argmax}_j \hat{r}_j$.} 

To evaluate forecast accuracy, we conduct a rolling window forecast exercise. %The response $y_t$ is generated according to the DGPs under the null hypothesis (Table \ref{SimDGP}). 
We use a  window of size $S=\lfloor 0.90 \cdot T\rfloor$. 
At each point $t=S, \ldots, T-1$, the  models are re-estimated and one-step-ahead forecasts are calculated. We evaluate the forecast accuracy of each selection-estimation technique combination by calculating the Mean Absolute Forecast Error\footnote{Similar conclusions can be drawn by looking at the Mean Squared Forecast Error.}
\begin{equation}
\text{MAFE} = \dfrac{1}{T-S}\sum_{t=S}^{T-1} \left | \hat{y}_{t+1} - y_{t+1}  \right |, \label{MAFEeq}
\end{equation}
where $\hat{y}_{t+1}$ is the predicted response for time $t+1$.
The MAFE is computed for each simulated time series, and their average over $N=100$ simulation runs is reported in Table \ref{MAFE}.
% In this forecast exercise, we simulate $N=100$ time series. We use a paired $t$-test  to test the null hypothesis that two competing methods have equal forecast accuracy.

\begin{table}[t]
\small
\caption{Average MAFE  for the four selection techniques (rows) and  four estimation techniques (columns). \label{MAFE}}
\resizebox{0.92\textwidth}{!}{\begin{minipage}{\textwidth}
\centering
\begin{tabular}{lllllcccc}
  \hline
Simulation design &&& Selection technique &&\multicolumn{4}{c}{Estimation technique} \\
&&&&&  ML & Adaptive Lasso  & Bayesian & Factor Model\\ 

  \hline
$T=100, k=25$ &&& All   && 0.093 & 0.089  & 0.116 & 0.129 \\
&&& Wald  test         && 0.082 & 0.082  & 0.121 & 0.086 \\
&&& Granger Lasso Selection && 0.089 & 0.085  & 0.118 & 0.121 \\
&&& Granger Lasso test && 0.082 & 0.082  & 0.120 & 0.086 \\
&&&&&&&&\\
$T=100, k=50$ &&& All   && 0.126 & 0.092  & 0.122 & 0.138 \\
&&& Wald  test            && 0.087 & 0.084  & 0.124 & 0.089 \\
&&& Granger Lasso Selection && 0.119 & 0.092  & 0.122 & 0.137 \\
&&& Granger Lasso test && 0.084 & 0.083  & 0.124 & 0.086 \\
&&&&&&&&\\
$T=100, k=75$ &&& All   && 0.208 & 0.089 & 0.123 & 0.141 \\
&&& Wald  test           && 0.117 & 0.088 & 0.121 & 0.107 \\
&&& Granger Lasso Selection && 0.170 & 0.091 & 0.123 & 0.140 \\
&&& Granger Lasso test && 0.083 & 0.080 & 0.119 & 0.085 \\
&&&&&&&&\\   
$T=40, k=150$ &&& All    && NA & 0.189 & 0.315 & 0.322  \\  
&&& Granger Lasso Selection && NA & 0.181 & 0.305 & 0.300 \\  
&&& Granger Lasso test && NA & 0.165 & 0.379 & 0.199 \\   \hline
\end{tabular}
\end{minipage} }
\end{table}

\smallskip

\textit{Results.}
Table \ref{MAFE} shows that 
% Granger Lasso Inference vs All
selecting predictor time series is better than taking all series, for all estimation techniques (except the Bayesian shrinkage estimator). %This result is in line with \cite{Bai08} who find  important gains in forecast accuracy from diffusion index models  by not using all predictors but by using fewer, informative predictors.
% Granger Lasso Inference vs Granger Lasso Selection
Among the selection techniques, improvements are larger with our Granger Lasso test compared to the Granger Lasso Selection approach.
Granger Lasso Selection  discards less blocks of time series compared to  the Granger Lasso test, yielding less parsimonious models and reduced forecast performance.
% Granger lasso Inference vs Standard Wald
When the number of time series increases relative to the length of the time series,  the Granger Lasso test also performs substantially better than the standard Wald test. Paired $t$-tests confirm that (in the majority of cases), the improvements  of the Granger Lasso test compared to the other selection techniques are significant. More precisely, the good performance of the Granger Lasso test is most pronounced in the high-dimensional designs: it performs significantly best - among the four selection techniques -  in 8 out of 12 cases (design $T=100, k=50$), 12 out of 12 cases (design $T=100, k=75$), and 6 out of 9 cases (design $T=40, k=150$).

For all simulation designs, the best forecast always involves the Granger Lasso test. Among the estimation techniques, the Adaptive Lasso performs best. After the first selection of predictive blocks of time series, the Adaptive Lasso can further reduce the number of predictor time series in the second step. This is most suited for settings with a few number of relevant predictor time series and a large number of irrelevant, noise predictor time series. Similar conclusions are obtained by \cite{Buhlmann10} who discuss a ``Twin Boosting" procedure for improved feature selection and prediction.

\section{The role of business and bank sentiment for macro-economic forecasting \label{application}}
We identify the most predictive industry segments for  future macro-economic developments using the Granger Lasso test from Section \ref{GC}. 

\subsection{Model \label{Model}}
We estimate 8  ARX models, one for each macro-economic indicator to predict. 
The time series $y_t$ entering  model \eqref{varp} is one of the 8 macro-economic indicators of Table \ref{Responses} taken in first differences. The vector  ${\bf x}_t$ contains the $k=150$ business and bank sentiment indicators in first differences at time $t$. We use differences to ensure stationarity of the time series.\footnote{Following standard practice, we first test for stationarity. A stationarity test of all individual time series using the Augmented Dickey-Fuller test indicates that most time series in levels are integrated of order 1.}
We estimate each ARX model using the Penalized Maximum Likelihood estimator from Section \ref{GC}. Then, we perform Granger Causality tests, one for each of the 20 blocks of sentiment indicators (cfr. Section \ref{data}). As such, we test if the opinion of a particular industry segment  - as measured through the Business Survey -  has incremental predictive power for the German macro-economic indicators. We repeat this exercise for each industry segment using the Bank Survey.

\subsection{Identifying the most predictive industries \label{GCResults}}
For each industry, Table \ref{BusinessBack} reports the $p$-value of the test that the opinion of that particular industry does not Granger Cause a particular macro-economic indicator. Significant results at the 1\% level are in bold.
We discuss the results by building on the sectoral classification framework %(e.g. \citealp{Kenessey87}) 
which distinguishes the primary, secondary, tertiary and quaternary sector.

% (1) primary sector (Agriculture, mining \& other industry), (2) secondary sector (Manufacturing and Construction), (3) tertiary sector (Wholesale, retail trade, transportation, food \& service), and (4) quaternary sector (remaining industry sectors).

\medskip

\begin{table}[t]
\caption{ \small $P$-values of the Granger Causality test with null hypothesis that the opinion of a particular industry segment (rows) %, as measured through the 9 indicators, 
does not Granger Cause a particular macro-economic indicator (columns). Significant results at the 1\% level are in bold. \label{BusinessBack}}
\resizebox{0.61\textwidth}{!}{\begin{minipage}{\textwidth}
\centering
\begin{tabular}{llllllllllll} \hline
&  &&& \multicolumn{8}{c}{Macro-economic indicators} \\ 
& Industry segment & Sector && IP-A1 & IP-A2 & IP-M & IP-E & IP-CaG & IP-CoG & RT & WS \\ 
  \hline
Business& Agriculture, mining \& other industry				& Primary &&  0.03 & 0.04  &  0.03 & 0.99  &  \textbf{0.01} & 0.01 & 0.01 & 0.84 \\
Survey & Manufacturing										& Secondary &&  \textbf{0.01} & 0.07  &  \textbf{0.00} & \textbf{0.00}  &  \textbf{0.00} & \textbf{0.01} & \textbf{0.00} & 0.37 \\
& Construction												& Secondary &&  \textbf{0.01} &  \textbf{0.00} &   0.01 & 0.04 &   \textbf{0.00} & 0.70 & \textbf{0.00} & 0.50 \\
& Wholesale, retail trade, transportation, food \& service	& Tertiary && 0.02 & \textbf{0.00}  &  0.04 & \textbf{0.01} &   0.02 & 0.923 &  0.27 & 0.06 \\
& Information \& communication								& Quaternary&&  0.92 & 0.02  &  0.90 & \textbf{0.00} &   0.02 & 0.50 & 0.04 & 0.04 \\
& Finance													& Quaternary&&  0.56 & 0.03  &  0.13 & \textbf{0.00} &   0.06&  0.04& 0.13& 0.39 \\
 & Real estate												& Quaternary&&  0.96& 0.84  &  0.26&  \textbf{0.01}  &  1.00 & \textbf{0.00} & \textbf{0.00}  &0.60 \\
& Administration \& support									& Quaternary&&  \textbf{0.01} & 0.03 &   \textbf{0.01} & \textbf{0.00} &   \textbf{0.00} & \textbf{0.01} & 0.21&  \textbf{0.00} \\
& Public services											& Quaternary&&   \textbf{0.00} & 0.02  &  0.23  &0.04 &   \textbf{0.00} &0.02 &0.86& 0.04 \\
& Other services											& Quaternary&& 0.05 & \textbf{0.00} &   \textbf{0.01}  &\textbf{0.00} &   \textbf{0.00}& 0.07 &0.66 &0.12 \\

&&&&&&&&&&& \\
Bank & Agriculture, mining \& other industry 				& Primary && 1.00 & 1.00 & 1.00 &  0.59 & 1.00 & 0.92 & 0.86 & 0.90 \\
Survey & Manufacturing 										& Secondary && 0.05 & 0.20 & 0.06 & 1.00 & 0.99 & 0.14 & 0.85 & 0.39 \\
& Construction 												& Secondary && 0.82 & 0.82 & 0.92 & \textbf{0.01}  & 1.00 & 0.70 & 0.84 & 0.03 \\
& Wholesale, retail trade, transportation, food \& service	& Tertiary && 1.00 & 0.76 & 0.98 & 1.00 &   \textbf{0.00}  & 0.04 & 0.53 & 0.23 \\
& Information \& communication								& Quaternary&& 0.72 & 0.02 & 0.09 & 1.00 &   0.04 & 0.53 & 0.05 & 0.79 \\
& Finance 													& Quaternary&& 0.98 & 1.00 & 1.00 & \textbf{0.01} &   1.00 & 0.40 & 0.09 & 0.08 \\
 & Real estate 												& Quaternary&& 0.76 & 0.90 & 0.60 & 1.00 &   1.00 & 0.73 & 0.80 & 0.62 \\
& Administration \& support 								& Quaternary && \textbf{0.01} & 0.29  &  \textbf{0.00} & 1.00 &   0.80 & 0.78 & 0.68 & \textbf{0.00} \\
& Public services 											& Quaternary&& 0.03 & 0.07 & \textbf{0.01} & 0.03 &   0.03 & 0.03 & 0.03 & 0.05 \\
& Other services 											& Quaternary&& 0.46 & 0.77 & 0.82 &  0.47 &    0.69 & 0.05 & 0.16 & 0.98 \\ \hline
\end{tabular}
\end{minipage}}
\end{table}

\textbf{Business Survey.}  
% Primary and Secondary Industry
The primary sector, unlike the other sectors, has almost no incremental predictive power. The primary sector's contribution to Germany's GDP is also the smallest. 
The secondary industry has most incremental predictive power for the macro-economic indicators to which these sectors contribute most (IP-A1, IP-A2, IP-M and IP-E).
% Tertiary Industry and Quaternary sector
Firms active in the tertiary and especially the quaternary sector have incremental predictive power for several macro-economic indicators.
This sector consists of the knowledge-based part of the economy, and accounts for roughly 65\% of Germany's GDP. Firms active in these sectors are at the heart of the whole economy.

\medskip

\textbf{Bank Survey.} The Bank Survey contains less incremental predictive power than the Business Survey. The predictive power of bank sentiment for predicting future macro-economic developments is  limited. This is in line with \cite{DellAriccia08} who find that the real effects of a banking crisis are limited in developed countries, in countries that have more access to foreign financing, and countries where banking crises are less severe, which all apply to Germany. 

\subsection{Robustness checks \label{GCRobustnessChecks}}
Our main research question is whether the sentiment of different industry segments has predictive power for  macro-economic indicators. Our methodology is also applicable to other ways of segmenting firms, as \textit{region} in which the are located  or according to their \textit{company size}. For our data, there are 10 regions and three company sizes. We re-estimate the 8 ARX models and perform the Granger Causality tests for the 20 regional blocks (i.e.\ 10 blocks for the Business Survey, 10 blocks for the Bank Survey). Likewise, we re-estimate the 8 ARX models and perform the Granger Causality tests for the 6 company size blocks (i.e.\ 3 blocks for the Business Survey, 3 blocks for the Bank Survey).

Similar as for the industry results discussed in Section \ref{GCResults}, we find that the business sentiment has more incremental predictive power compared to the bank sentiment.
Furthermore, Germany's largest geo-economical regions, Ruhr area and the Southern states, have most incremental predictive power for the macro-economic indicators to which their day-to-day business contributes most, i.e.\ IP-A1, IP-A2, IP-M,  IP-E and IP-CaGo, IP-CoGo respectively. 
Finally, small- and medium-sized companies have more incremental predictive power than large companies. Germany is dominated by small- to medium-sized companies who are global market leaders in their segments, and, hence, those might be best at evaluating Germany's economy. Detailed results are available from the authors upon request.
%Classifications of business segments according to region and industry might distinguish differences in know-how between businesses more outspokenly than a classification based on company size. These differences in know-how might result in these segments being most predictive for those macro-economic indicators that are most closely tied to this know-how. 

\section{Forecasting German macro-economic developments \label{forecast}}
%We investigate the predictive power of the sentiment indicators for forecasting future macro-economic developments. 
We perform a rolling-window forecast exercise  using a  window of length  $S=30$. For each time window, we estimate the 8 ARX models.  We use the same selection and estimation techniques as in Section \ref{SimForecast}, except for the standard Wald test and the ML estimator which are not available since the number of time series exceeds the time series length. Next, one-step-ahead forecasts are computed for $t=S+1,\ldots,T$.
We report the Mean Absolute Forecast Error, see equation \eqref{MAFEeq}, for each macro-economic indicator and each selection-estimation technique combination in Table \ref{MAFEIndustry}. 

\begin{table}[t]
\small
\caption{Mean Absolute Forecast Error for the three selection techniques (rows), the three estimation techniques (columns), and the 8 macro-economic indicators (blocks).  \label{MAFEIndustry}}
\resizebox{0.70\textwidth}{!}{\begin{minipage}{\textwidth}
\centering
\begin{tabular}{llccc|lccc} \hline
Selection technique & Response &\multicolumn{3}{c}{Estimation technique} & Response& \multicolumn{3}{c}{Estimation technique} \\
&&  Adaptive Lasso & Bayesian  & Factor Model   & & Adaptive Lasso   & Bayesian  & Factor Model  \\ \hline
%&&&&&&&&&&&& \\

All & IP-A1               & 1.460  & 0.921 &  1.275 & IP-CaGo & 2.734 & 1.892 & 3.147	\\
Granger Lasso Selection &         & 1.460  & 0.921& 1.275 	&& 2.734  & 1.892 & 3.147    \\
Granger Lasso test  &        & 1.138  & 0.962 & 0.937  	&& 3.707  & 1.834 & 2.926    \\
&&&&&&& & \\
All & IP-A2              & 1.462  & 0.817 & 1.207 & IP-CoGo & 1.142  & 0.609 & 0.918 	\\
Granger Lasso Selection  &        & 1.462  & 0.817 & 1.207  	&& 1.142  & 0.609 & 0.918     \\
Granger Lasso test  &        & 0.567 & 0.640 & 1.006 	&& 0.777  & 0.617 & 0.915    \\
&&&&&&& & \\

All & IP-M               & 1.720 & 1.117 & 1.641& RT & 2.025  & 1.109 & 1.723 	\\
Granger Lasso Selection  &        & 1.720  & 1.117 & 1.641  	&& 2.025  & 1.109 & 1.723     \\
Granger Lasso test  &        & 1.688  & 1.090 & 1.342	&& 1.140 & 1.035 & 1.510  \\
&&&&&&&&  \\

All & IP-E                & 2.237  & 1.171 & 2.105 & WS & 1.524  & 0.530 & 0.800 	\\
Granger Lasso Selection  &        & 2.237  & 1.171 & 2.105 	&& 1.524  & 0.530 & 0.800       \\
Granger Lasso test &         & 1.249  & 0.959 & 1.601	&& 0.566  & 0.685 & 0.677    \\
 \hline
\end{tabular}
\end{minipage}}
\end{table}

%We indicate in bold the MAFE for the best performing selection technique, for each macro-economic indicator and estimation technique.
Among the selection techniques, the proposed Granger Lasso test performs best. It attains the lowest value of the MAFE in 20 out of 24 cases (84\% of the cases). 
The MAFEs when either all industries are used or when  Granger Lasso Selection is used are close to each other. It turns out   that the latter (overall) does not discard any of the industry blocks. In contrast, a much more parsimonious model is obtained using the Granger Lasso test. These parsimonious models lead to an improved forecast accuracy, in the majority of cases.

For the Adaptive Lasso, the Granger Lasso test leads to the lowest MAFE for 7 out of 8 macro-economic indicators. The MAFEs with the Granger Lasso test are, on average, 40\% lower compared to the other selection techniques. After the first selection step where either an entire block of business or bank sentiment indicators is selected or not, the Adaptive Lasso  allows some of the time series belonging to a one of the selected blocks to be discarded in this second stage. Further reducing the number of relevant predictor time series within the selected blocks improves forecast accuracy.

In line with the results of our simulation study, pre-selecting based on the Granger Lasso test is less favorable for the Bayesian shrinkage estimator compared to the other estimation techniques. Nevertheless, the Granger Lasso test in combination with the Bayesian shrinkage estimator still leads to the lowest MAFE for 5 out of 8 macro-economic indicators, with an average reduction in MAFE of 10\%.

For the Factor Model, the Granger Lasso test consistently leads to the lowest MAFE. The MAFEs with the Granger Lasso test are, on average, 20\% lower compared to the other selection techniques. Discarding the least predictive industry blocks in this high-dimensional data set and estimating the factors based on the most predictive industry blocks thus leads to important gains in forecast accuracy. 
This result is in line with \cite{Bai08} who find  important gains in forecast accuracy from diffusion index models  by not using all predictors but by using fewer, informative predictors. 

\medskip

\textbf{Robustness checks.} We investigate the robustness of the results to the choice of segmentation criterion. We repeat the same forecast exercise using the region segments and company size segments instead of the industry segments (cfr. Section \ref{GCRobustnessChecks}). The conclusions obtained with either the industry, region or company size segments are very similar.
For the regional segments, the Granger Lasso test is the best performing selection technique and attains the lowest value of the MAFE in 71\% of the cases (17 out of 24). Similarly for the company size segments where the Granger Lasso test leads towards the lowest MAFE in 71\% of the cases (17 out of 24). Detailed results are available from the authors upon request.

\section{Discussion \label{conclusion}}
% Main contribution
This paper presents a high-dimensional Granger Causality test. It detects the most predictive industry segments for future macro-economic developments. For this purpose, we use both business and  bank sentiment surveys answered by firms across Germany. Not all industry-specific sentiment indicators are equally predictive for all macro-economic indicators.  Industries contain most predictive power for the macro-economic indicators most closely tied to their day-to-day business activities.

%% Alternative approaches 
%We combine all 150 sentiment indicators in one high-dimensional data set. Other approaches are possible.
%One could compute an aggregate indicator for each block  and then test the significance of this aggregate indicator. This approach has two disadvantages. First, by aggregating, information is lost. Second, weights for the disaggregate indicators need to be determined to calculate the aggregate indicator, which involves subjectivity. Therefore, we opt to perform joint hypothesis tests using the disaggregate sentiment indicators.
%Another approach would be to estimate different models, one for each region or industry segment like \cite{Martinsen14}. However, when the number segments is large, a large number of models needs to be estimated. 

% Selecting informative business segments
 Our forecast exercise shows that important gains in forecast accuracy can be obtained by not using all industry segments, but by first selecting the most predictive ones using the Granger Lasso test. This selection of the most pertinent industry segments provides important information for institutes conducting these sentiment surveys. For instance, instead of equally spreading respondents among all segments, the number of respondents in predictive segments could be increased, whereas the number of respondents in non-predictive segments could be decreased. Alternatively, non-predictive segments could even be completely discarded, which provides an opportunity to obtain cost savings.

% Consumer Survey National Bank of Belgium
The identification of pertinent respondents also applies to consumer sentiment surveys. In the large literature on consumer sentiment, this topic has received little attention. We perform a similar exercise as described in this paper using a consumer sentiment survey data set from the National Bank of Belgium. Sentiment indicators are available for different classes of consumers' net disposable income, profession, employment status, education, age and gender. We study their predictive power for several retail trade indicators. The profession, education, and age sentiment indicators contain most predictive power. Again, important gains in forecast accuracy can be obtained  by first selecting the most predictive sentiment indicators (for a specific target variable of interest) instead of using all indicators.

% Importance bank confidence countries hit by financial crisis.
In our sentiment application, the Business Survey contains more predictive power than the Bank Survey. Future research could further deepen our understanding on the usefulness of bank sentiment. It would be interesting to investigate if this sentiment differs between, for instance, countries that are more or less severely hit by banking crises, and developed or developing countries. The study of sentiment with respect to the banking sector opens a rich area of new research on sentiment surveys.

\paragraph{Acknowledgments} The authors gratefully acknowledge financial support from the FWO (Research Foundation Flanders, contract number 11N9913N). We would also like to thank EUWIFO for providing the data.

\bigskip

%\newpage

% References
\bibliographystyle{asa}
\bibliography{GCref}

\end{document}